\documentclass[aps,prl,floatfix,twocolumn,showpacs,preprintnumbers,amsmath,amssymb]{revtex4-1}%

\usepackage{graphicx}
\usepackage{dcolumn}
\usepackage{bm}

\begin{document}

\preprint{APS}

\title{Crossover from fast relaxation to physical aging in colloidal adsorption at fluid interfaces}
\author{Carlos E. Colosqui}
\affiliation{Benjamin Levich Institute, 
City College of the City University of New York, New York, NY 10031, USA}
\author{Jeffrey F. Morris}
\affiliation{Benjamin Levich Institute, 
City College of the City University of New York, New York, NY 10031, USA}
\author{Joel Koplik}
\email{koplik@sci.ccny.cuny.edu} 
\affiliation{Benjamin Levich Institute, 
City College of the City University of New York, New York, NY 10031, USA}
\date{}
%
%%%%%%%%%%%%%%%%%%%%%%%%%%%%%%%%%%%%%%%%%%%%%%%%%%%%%%%%%%%%%%%%%%%%%%%%%%%%%%%%%%%%%%%%%%%%%%%%%%%
%
% ABSTRACT
%
%%%%%%%%%%%%%%%%%%%%%%%%%%%%%%%%%%%%%%%%%%%%%%%%%%%%%%%%%%%%%%%%%%%%%%%%%%
%
\begin{abstract}
The adsorption dynamics of a colloidal particle at a fluid interface is studied theoretically and numerically, documenting distinctly different relaxation regimes.
The adsorption of a perfectly smooth particle is characterized by a fast exponential relaxation to thermodynamic equilibrium where the interfacial free energy has a minimum. 
The short relaxation time is given by the ratio of viscous damping to capillary forces.
Physical and/or chemical heterogeneities in a colloidal system, however, can result in multiple minima of the free energy giving rise to metastability.
In the presence of metastable states we observe a crossover to a slow logarithmic relaxation reminiscent of physical aging in glassy systems.
The long relaxation time is determined by the thermally-activated escape rate from metastable states.
Analytical expressions derived in this work yield quantitative agreement with molecular dynamics simulations and recent experimental observations.
This work provides new insights on the adsorption dynamics of colloidal particles at fluid interfaces. 
\end{abstract}
%
%\keywords{}
\maketitle
%
%
%%%%%%%%%%%%%%%%%%%%%%%%%%%%%%%%
%   INTRODUCTION
%%%%%%%%%%%%%%%%%%%%%%%%%%%%%%%%
%
The adsorption and binding of colloidal particles to fluid interfaces is relevant to numerous natural and industrial processes.
Novel technological applications in areas that range from materials science to renewable energy \cite{particles,velev2009,*goesmann2010}, and from food science to biomedicine \cite{mezzenga2005,dinsmore2002,*velikov2008} demand advancements in the fundamental understanding of colloidal adsorption dynamics.
Standard models based on continuum thermodynamics \cite{particles,pieranski1980} predict a monotonic relaxation to an equilibrium position where the contact angle with the interface is given by Young's law \cite{israelachvili1989,*bormashenko2012}. 
According to standard models, this equilibrium position corresponds to a stable state determined by the (global) minimum of the Helmholtz free energy.
For micrometer- or nanometer-size particles, the energy decrease at the equilibrium state can be orders of magnitude larger than the thermal energy, and thus strong interfacial forces are expected to cause a spontaneous adsorption with rapid relaxation to equilibrium.

Nevertheless, the dynamics of colloidal adsorption remains poorly understood for systems of great practical interest (e.g. functionalized colloidal particles). 
Fundamental issues arise when the equilibrium contact angle is difficult to determine, e.g. due to the presence of physical and/or chemical heterogeneities \cite{israelachvili1989,*bormashenko2012,paunov2003,*arnaudov2009}.
Furthermore, even when the equilibrium contact angle can be properly determined it is frequently observed that the adsorption of a colloidal particle is neither fast nor spontaneous, and requires some form of external actuation (mechanical, thermal, or chemical) to be initiated \cite{stocco2011,du2010,kaz2011}. 
Notably, recent experimental work \cite{kaz2011} reported a slow logarithmic relaxation to equilibrium after initiating the adsorption of a micrometer-size particle at a water-oil interface. 
The unexpected observation was attributed to nanoscale surface heterogeneities \cite{kaz2011}.
A logarithmic relaxation is reminiscent of physical aging in glassy systems having complex free energy landscapes and metastability \cite{Strum1977}. 

These phenomena suggest that the presence of energy barriers associated with microscale heterogeneities should be considered in order to describe the adsorption dynamics of colloidal particles.
In this Letter, we study the dynamics of adsorption in the presence of multiple metastable states caused by spatial fluctuations in the interfacial free energy.
We propose a model that quantitatively describes recent experimental observations \cite{kaz2011}.
Our derived analytical expressions and numerical simulations  reveal a non-trivial adsorption dynamics with crossovers from logarithmic to exponential relaxation.

\begin{figure}
\center
\includegraphics[angle=0,width=1.0\linewidth]{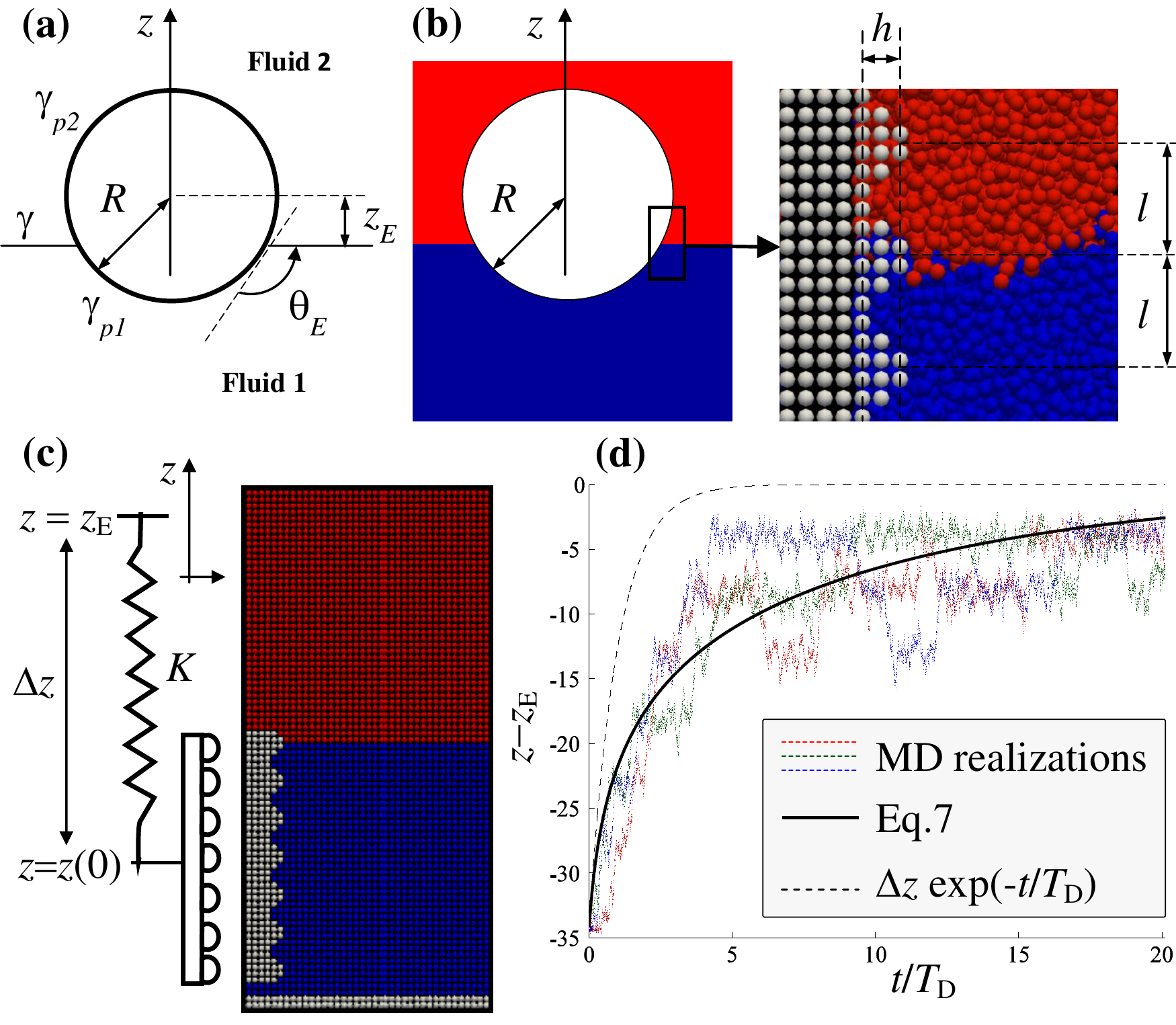}
\caption{Problem description and modeled system.
(a) Colloidal particle of radius $R$ straddling a sharp fluid interface.
(b) Particle with modeled microscale heterogeneities of size $l \ll R$.
Interfacial energy fluctuations have wavelength $l$ and amplitude $\Delta U=\gamma h w$ associated with surface roughness of height $h$ and width $w\sim h$.
The system free energy is given by Eq.~\ref{eq:energy}.
(c) Simplified system for MD simulations with free energy given by Eq.~\ref{eq:energy}. 
A linear ``spring'' force $F=K \Delta z$ drives the relaxation to equilibrium at $z=z_E$.
(d) Trajectories $z(t)$ from MD simulations (dotted lines) for different realizations ($L_H$=$Kl/2k_BT$=7.4, $\Delta U/k_BT$=4, and $\Delta z$=-4.5$L_H$). 
Analytical expressions are plotted for comparison (see legend).}
\label{fig1}
\end{figure}

%%%%%%%%%%%%%%%%%%%%%%%%%%%%%%%%%%%%%%%%%%%%%%%%%%%%%%%%%%%%%%%%%%%%%%%%%%
%   THEORY: SMOOTH SPHERICAL PARTICLE
%%%%%%%%%%%%%%%%%%%%%%%%%%%%%%%%%%%%%%%%%%%%%%%%%%%%%%%%%%%%%%%%%%%%%%%%%%
We begin our analysis with the equation of motion for a colloidal particle that straddles the interface between two fluids as illustrated in Fig.~\ref{fig1}(a).
Assuming that the particle undergoes Brownian motion, and neglecting the action of external fields, its vertical position $z$ is governed by a Langevin equation
\begin{equation}
m \ddot{z}=\sqrt{2 k_B T \xi} \eta(t)-\xi\dot{z}+F(z); 
\label{eq:langevin}
\end{equation}
here $m$ is the particle mass, $k_B T$ is the thermal energy of the surrounding fluids, $\eta(t)$ is a zero-mean unit-variance Gaussian noise, $\xi$ is the viscous friction coefficient, and $F(z)=-\partial U/\partial z$ is the interfacial or {\it capillary} force determined by the interfacial free energy $U$.
%
%%%%%%%%%%%%%%%%%%%%%%%%%%%%%%%%%%%%%%%%%%%%%%%%%%%%%%%%%%%%%%%%%%%%%%%%%%
%   THEORY: SMOOTH SPHERICAL PARTICLE
%%%%%%%%%%%%%%%%%%%%%%%%%%%%%%%%%%%%%%%%%%%%%%%%%%%%%%%%%%%%%%%%%%%%%%%%%%
%
In the framework of continuum thermodynamics, the energy to form an interface is the product of the interfacial area and the corresponding surface tensions: $\gamma$ for the fluid-fluid interface; $\gamma_{p1}$ and $\gamma_{p2}$ for the interfaces between the particle and each fluid phase.
Hence, for a spherical particle of radius $R$ with its center at vertical position $z$ and a sharp fluid interface located at $z=0$, the interfacial free energy can be cast as \cite{pieranski1980} 
\begin{equation}
U_S(z)=\textstyle{\frac{1}{2}} K (z-z_E)^2-C~~\mathrm{for}~~|z|\le R,
\label{eq:energy_smooth} 
\end{equation}
where $z_E$ is the equilibrium position, while $K=2 \pi \gamma$ and $C=\pi \gamma  (R-z_E)^2$ are positive constants. 
According to Eq.~\ref{eq:energy_smooth} the interfacial force is linear, $F=-K (z-z_E)$. 
Employing this linear force and neglecting small inertial effects, the solution of Eq.~\ref{eq:langevin} gives an average displacement $<z>=z_E+\Delta z\exp(-t/T_D)$, where $\Delta z=z(0)-z_E$ is the distance from equilibrium at $t=0$ and $T_D=\xi/K$ is the viscous decay time.
For reference, we must note that the decay time to equilibrium is $T_D\sim$~0.1$\mu$s for a one-micron radius particle adsorbed at a water-oil interface.
Thus, the standard model predicts a fast exponential relaxation to equilibrium for a perfectly smooth particle. 
%

%%%%%%%%%%%%%%%%%%%%%%%%%%%%%%%%%%%%%%%%%%%%%%%%%%
% THE CONTACT ANGLE & PARTICLE POSITIONS
%%%%%%%%%%%%%%%%%%%%%%%%%%%%%%%%%%%%%%%%%%%%%%%%%%
A few comments are in order.
For the idealized case of a perfectly smooth and spherical particle that straddles a flat interface, there is a one-to-one correspondence between the particle position $z$ and the observed contact angle $\theta=\mathrm{acos}(-z/R)$ which is constant along the circular contact line located at $z=0$.
According to Young's law for the equilibrium contact angle, $\cos\theta_E=(\gamma_{p2}-\gamma_{p1})/\gamma$, the particle will straddle the two fluids at an equilibrium position $z_E= - R \cos \theta_E$ when $|\gamma_{p2}-\gamma_{p1}|<\gamma$.
%
%%%%%%%%%%%%%%%%%%%%%%%%%%%%%%%%%%%%%%%%%%%%%%%%%%%%%%%%%%%%%%%%%%%%%%%%%%
%  LINE TENSION & HETEROGENEITIES
%%%%%%%%%%%%%%%%%%%%%%%%%%%%%%%%%%%%%%%%%%%%%%%%%%%%%%%%%%%%%%%%%%%%%%%%%%
Both Young's law and Eq.~\ref{eq:energy_smooth} for $U_S(z)$ apply to homogeneous and regular contact lines.
In order to model heterogeneities that cause fluctuations of the contact line shape and its mean position, we will neglect the line tension term used in alternative models \cite{aveyard1996,marmur1998} and introduce spatial fluctuations in the interfacial free energy. 
%   

%%%%%%%%%%%%%%%%%%%%%%%%%%%%%%%%%%%%%%%%%%%%%%%%%%%%%%%%%%%%%%%%%%%%%%%%%%%%%%%%%%%%%%%%%%%%%%%%%%%%%%%%%%%%%%%%%%%%%%%%%%%%%%%%
% Figure1: Problem setup
%%%%%%%%%%%%%%%%%%%%%%%%%%%%%%%%%%%%%%%%%%%%%%%%%%%%%%%%%%%%%%%%%%%%%%%%%%%%%%%%%%%%%%%%%%%%%%%%%%%%%%%%%%%%%%%%%%%%%%%%%%%%%%%%
%%%%%%%%%%%%%%%%%%%%%%%%%%%%%%%%%%%%%%%%%%%%%%%%%%%%%%%%%%%%%%%%%%%%%%%%%%%%%%%%%%%%%%%%%%%%%%%%%%%%%%%%%%%%%%%%%%%%%%%%%%%%%%%%
%

%%%%%%%%%%%%%%%%%%%%%%%%%%%%%%%%%%%%%%%%%%%%%%%%%%%%%%%%%%%%%%%%%%%%%%%%%%
%  THEORY: FREE ENERGY W/ FLUCTUATIONS
%%%%%%%%%%%%%%%%%%%%%%%%%%%%%%%%%%%%%%%%%%%%%%%%%%%%%%%%%%%%%%%%%%%%%%%%%%
We are interested in a simple analytical description of microscale physical/chemical heterogeneities, as illustrated in Fig.~\ref{fig1}(b), that can give rise to significant energy barriers $\Delta U>k_BT$.
For this purpose, we introduce spatial fluctuations of the interfacial free energy with the general form 
$\textstyle{\frac{1}{2}}\Delta U \sin(\lambda \theta+\phi)$; 
here $\textstyle{\frac{1}{2}} \Delta U$ is the amplitude of the fluctuation, 
$l=2\pi/\lambda$ is its wavelength, 
and $\phi$ is an arbitrary phase.  
For the sake of analytical tractability we will consider a single-mode fluctuation of small wavelength, $l \ll R$, equilibrium contact angles near neutral wetting, $70^\circ \lesssim \theta_E \lesssim 110^\circ$, and conditions where the particle center is close to the fluid interface, $|z/R|\ll 1$.
Curvature effects, of order ${\cal O}((z/R)^2)$, can thus be neglected and the free energy is expressed as
\begin{equation}
U(z)=U_S(z)+ \textstyle{\frac{1}{2}} \Delta U \sin(\lambda z+\phi)
\label{eq:energy}
\end{equation}
while the capillary force on the particle is
\begin{equation}
F(z)=- K (z-z_E) - \textstyle{\frac{1}{2}} \lambda \Delta U \cos(\lambda z+\phi).
\label{eq:force}
\end{equation}

A simplified system with dynamics governed by Eq.~\ref{eq:langevin} and free energy given by Eq.~\ref{eq:energy} is depicted in Fig.~\ref{fig1}(c).
The simplified system consists of a solid surface element with sinusoidal roughness (wavelength $l$, height $h$, and width $w$) that is in contact with the fluid interface, and thus produces interfacial energy fluctuations of magnitude $\Delta U=\gamma h w$ as it undergoes Brownian motion.
Just as if the surface element formed part of a much larger spherical particle, a linear force drives the system toward equilibrium $z\to z_E$; this force corresponds to the free energy contribution $U_S$ in Eq.~\ref{eq:energy_smooth} caused by the macroscopic curvature of the particle.   
The decomposition of macroscale and microscale features makes the dynamics tractable via molecular dynamics (MD) simulation \footnote{MD simulations required extremely long runs (i.e. over 10 million timesteps) and several (7--10) independent realizations; running in parallel (7 CPUs at 2.20GHz) the simulations required over 300 hours of CPU time.}.
Simulated trajectories $z(t)$ from individual MD realizations shown in Fig.~\ref{fig1}(d) exhibit several metastable states with a lifetime that increases as the equilibrium position $z_E$ is approached.

%%%%%%%%%%%%%%%%%%%%%%%%%%%%%%%%%%%%%%%%%%%%%%%%%%%%%%%%%%%%%%%%%%%%%%%%%%%
%  ESCAPE FROM METASTABLE STATES
%%%%%%%%%%%%%%%%%%%%%%%%%%%%%%%%%%%%%%%%%%%%%%%%%%%%%%%%%%%%%%%%%%%%%%%%%%%
According to Eqs.~\ref{eq:energy}--\ref{eq:force} there are multiple minima in the free energy for 
$|z-z_E|< \pi \Delta U/K l$, 
and sufficiently close to equilibrium the system must exhibit metastability.
The basins of attraction of each metastable state are centered at the local minima 
$z_{o}=z_E +l(n-\textstyle{\frac{1}{4}}-\phi/2\pi)+{\cal O}(\epsilon)$ and 
each basin is bounded by two neighboring maxima at 
$z_{\pm}= z_o\pm \textstyle{\frac{1}{2}} l+{\cal O}(\epsilon)$; 
here $n$ is any integer and 
$\epsilon=Kl|z-z_E|/ \pi \Delta U$ 
is a small parameter.
%
%%%%%%%%%%%%%%%%%%%%%%%%%%%%%%%%%%%%%%%%%%%%%%%%%%%%%%%%%%%%%%%%%%%%%%%%%%
%  THEORY: FREE ENERGY W/ FLUCTUATIONS
%%%%%%%%%%%%%%%%%%%%%%%%%%%%%%%%%%%%%%%%%%%%%%%%%%%%%%%%%%%%%%%%%%%%%%%%%%
A particle undergoing Brownian motion (see Fig.~\ref{fig1}(d)) will transition, or ``hop'' back and forth, between metastable states at a local rate \cite{kramers}
\begin{equation}
\Gamma_{\pm}(z)
=\frac{1}{2\pi\xi}
\sqrt{\frac{\partial^2 U(z_o)}{\partial z^2} \left|\frac{\partial^2 U(z_\pm)}{\partial z^2}\right|} 
\exp\left(-\frac{\Delta U_{\pm}}{k_B T}\right)
\label{eq:kramers}
\end{equation}
predicted for $|z-z_o|<\textstyle{\frac{1}{2}} l$.
The energy barriers, $\Delta U_{\pm}=U(z_\pm)-U(z_o)$, in the forward/backward direction determine the Arrhenius exponential factor.
The prefactor employed in Eq.~\ref{eq:kramers} is valid for overdamped systems \cite{kramers,hanggi1986} where $\xi>\sqrt{m \left|\partial^2 U(z_\pm)/\partial z^2\right|}$ (here $m$ is the particle mass).
For $|z-z_E| \ll \pi \Delta U/K l$ the motion is dominated by thermally-activated hopping, and the ensemble-averaged speed $<\dot{z}>$ of the particle is determined by an ordinary differential equation
\begin{equation}
<\dot{z}>=\textstyle{\frac{1}{2}} l (\Gamma_+-\Gamma_-).
\label{eq:motion}
\end{equation}
%
%Similar equations of motion have been employed to model the dynamics of contact lines \cite{haynes1969,*blake2006}.
%
We solve analytically the equation of motion in Eq.~\ref{eq:motion} to obtain the average trajectory 
\begin{equation}
\left< z \right>= z_E+
L_H \log \left[ \frac{1+A_H \exp(-t/T_H)}{1-A_H \exp(-t/T_H)}\right],
\label{eq:deltaz}
\end{equation}
which depends on three independent parameters:
the characteristic hop length $L_H=2 k_B T/ K l$;
the trajectory ``amplitude'' $A_H=\tanh(\textstyle{\frac{1}{2}}\Delta z/L_H)$ determined by the initial separation from equilibrium $\Delta z=z(0)-z_E$ at time $t=0$;
and the characteristic hop time 
\begin{equation}
T_H= T_D 
\left(\frac{L_H}{l}\right) 
\frac{2\pi}{\sqrt{|\Phi^2-1|}}
\exp\left(\frac{\Delta U}{k_BT}+\frac{1}{4}\frac{l}{L_H}\right).
\label{eq:time}
\end{equation}
In Eq.~\ref{eq:time} we introduced the ratio $\Phi= {\textstyle\frac{1}{2}}\Delta U \lambda^2/K$ between the free energy curvature of the modeled sinusoidal fluctuation and that of a smooth spherical particle.  
The characteristic hop length, $L_H$, and time, $T_H$, in our model are thus determined by two independent parameters, (i) the dimensionless energy barrier $\Delta U/k_BT$ and (ii) the dimensionless wavelength $l/\sqrt{k_BT/\gamma}$, while the dynamics of relaxation also depends on the dimensionless initial condition $\Delta z/L_H$. 
%
%%%%%%%%%%%%%%%%%%%%%%%%%%%%%%%%%%%%%%%%%%%%%%%%%%%%%%%%%%%%%%%%%%%%%%%%%%
%   LOG-EXP LIMIT OF THE HOPPING REGIME
%%%%%%%%%%%%%%%%%%%%%%%%%%%%%%%%%%%%%%%%%%%%%%%%%%%%%%%%%%%%%%%%%%%%%%%%%%
Sufficiently far from equilibrium when $|z-z_E|/L_H\gg 1$, we have 
$\left<z\right>=z_E+L_H \log[{\textstyle \frac{1}{2}} t/T_H+\exp(\Delta z/L_H)]$, 
which is equivalent in form to the logarithmic expression recently employed in Ref.~\cite{kaz2011} to fit experimental observations by treating $L_H$ and $T_H$ as adjustable parameters.
It is noteworthy that Eq.~\ref{eq:deltaz} predicts a crossover to exponential relaxation, 
$\left<z\right>=z_E+\Delta z \exp(-t/T_H)$, for $|z-z_E|/L_H\ll 1$ 
when the particle is very close to equilibrium.
%

%
%
%%%%%%%%%%%%%%%%%%%%%%%%%%%%%%%%%%%%%%%%%%%%%%%%%%%%%%%%%%%%%%%%%%%%%%%%%%%%%%%%%%%%%%%%%%%%%%%%%%%%%%%%%%%%%%%%%%%%%%%%%%%%%%%%
% Figure2: MD vs. LD vs theory
%%%%%%%%%%%%%%%%%%%%%%%%%%%%%%%%%%%%%%%%%%%%%%%%%%%%%%%%%%%%%%%%%%%%%%%%%%%%%%%%%%%%%%%%%%%%%%%%%%%%%%%%%%%%%%%%%%%%%%%%%%%%%%%%
\begin{figure}
\center
\includegraphics[angle=0,width=1.0\linewidth]{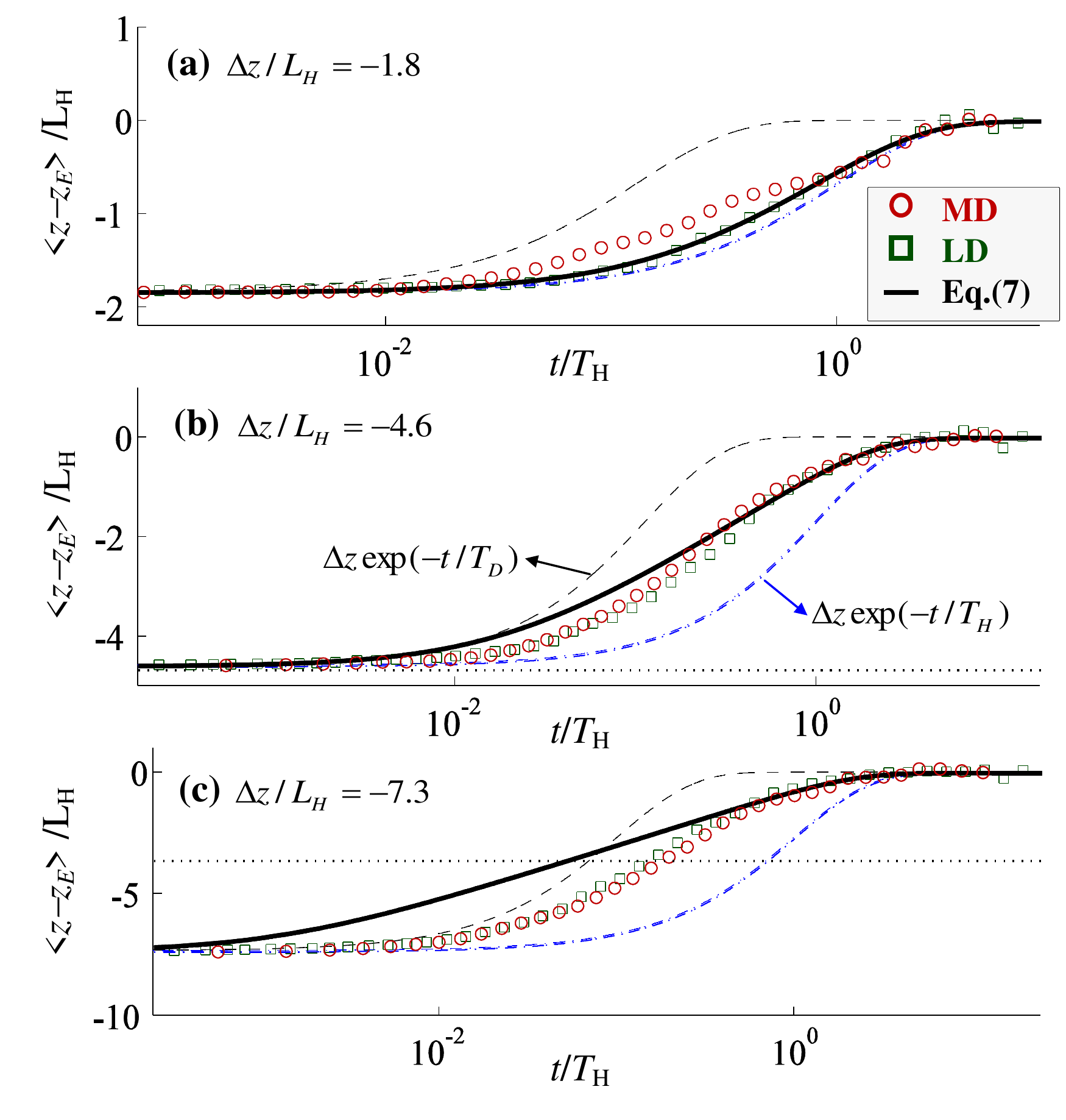}
\caption{Mean relaxation trajectories $\left<z(t)-z_E\right>$.
Horizontal dotted lines indicate the distance from equilibrium $\Delta z_H={\textstyle \frac{1}{2}} \pi \Delta U/ l$ above which hopping dominates.
Markers indicate numerical simulations via MD and LD (see legend).
Solid lines indicate analytical predictions from Eq.~\ref{eq:deltaz}.
Dashed lines indicate exponential relaxations at the viscous decay time $T_D=\xi/K$ and the hop time $T_H$ from Eq.~\ref{eq:time} (see labels).
The energy barrier is $\Delta U=4 k_BT$, and the fluctuation wavelength is $l=4.5 \sqrt{k_BT/\gamma}$ for all three initial conditions: 
(a)$\Delta z/L_H$=-1.8;
(b)$\Delta z/L_H$=-4.6; and
(c)$\Delta z/L_H$=-7.3.
}
\label{fig:md-ld-th}
\end{figure}
%%%%%%%%%%%%%%%%%%%%%%%%%%%%%%%%%%%%%%%%%%%%%%%%%%%%%%%%%%%%%%%%%%%%%%%%%%%%%%%%%%%%%%%%%%%%%%%%%%%%%%%%%%%%%%%%%%%%%%%%%%%%%%%%
%
%
%%%%%%%%%%%%%%%%%%%%%%%%%%%%%%%%%%%%%%%%%%%%%%%%%%%%%%%
% NUMERICAL METHODS
%%%%%%%%%%%%%%%%%%%%%%%%%%%%%%%%%%%%%%%%%%%%%%%%%%%%%%%
To verify our analytical predictions, derived under the assumption of a quasi-static process and (overdamped) Brownian motion \cite{kramers,hanggi1986}, we perform numerical simulations via standard MD techniques \cite{rapaport}. 
The simulated system, illustrated in Fig.\ref{fig1}(c), has three atomic species: ($i$=1) fluid 1; ($i=2$) fluid 2; ($i=3$) the solid particle and bottom wall.
As in previous work \cite{koplik2006,*drazer2002}, our MD simulations employ generalized Lennard-Jones potentials 
$V(r)=4\epsilon[(r/\sigma)^{-12}-c_{ij}(r/\sigma)^{-6}]$,
where $\epsilon$ is the interaction energy,
$\sigma$ is roughly the atomic radius,
$r$ is the distance between any two atoms, 
and $c_{ij}=c_{ji}$ is the interaction coefficient between species ($i,j$ = 1--3).
In this work we set $c_{ii}=1$ for self interactions, while in the case of cross interactions we set $c_{12}=0.5$ for the fluids, $c_{13}=c_{23}=0.35$ for the fluids and Brownian particle, and $c_{13}=c_{23}=0.8$ for the fluids and stationary bottom wall.
At a simulated constant temperature $T=3 \epsilon/k_B$, maintained by a Nos\'{e}-Hoover thermostat, and a mass density $\rho=0.8/\sigma^3$, the fluids are macroscopically immiscible and the surface tension measured across a plane interface \cite{koplik1995} is $\gamma\simeq 1.4 k_B T/\sigma^2$.
Solid surfaces exhibit neutral wetting ($\theta_E=90^\circ$) given the symmetry of fluid-solid interactions. 
All atoms have a unit mass and are initialized on a fcc lattice (cf. Fig.~\ref{fig1}(c)) with spatial spacing $\Delta x=\sqrt[3]{1/\rho}$.
The solid particle and wall, carved from the fcc lattice, are neutrally buoyant.
The particle has length $L_T=40\Delta x$ and a sinusoidal roughness with wavelength $l=6\Delta x$, height $h=2\Delta x$, and width $w=2h$, which in turn produces surface energy fluctuations of amplitude $\Delta U=\gamma h w$.    
In different numerical realizations, fluid atoms are initialized with random velocities and the particle is allowed to move in the vertical $z$ direction after thermal equilibrium is attained.
The hop length $L_H=2 k_B T/ K l$ in MD simulations is modified by adjusting the ``spring'' stiffness K. 
In addition, we perform Langevin dynamics (LD) simulations which are equivalent to the numerical solution of Eq.~\ref{eq:langevin}.

%%%%%%%%%%%%%%%%%%%%%%%%%%%%%%%%%%%%%%%%%%%%%%%%%%%%%%%%%%%%%%%%%%%%%
% RESULTS DESCRIPTION
%%%%%%%%%%%%%%%%%%%%%%%%%%%%%%%%%%%%%%%%%%%%%%%%%%%%%%%%%%%%%%%%%%%%%
Ensemble-averaged trajectories $\left<z(t)\right>$ are reported in Fig.~\ref{fig:md-ld-th} for three different displacement amplitudes $|\Delta z|=|z(0)-z_E|\simeq$ 2--7$L_H$. 
The simulated case corresponds to an energy barrier $\Delta U=4 k_BT$ and a fluctuation wavelength $l=4.5 \sqrt{k_BT/\gamma}$.
Very close to equilibrium for $|\Delta z|/L_H<2$ (cf. Fig.~\ref{fig:md-ld-th}(a)), the relaxation is exponential at the slower rate $1/T_H$ predicted by Eq.~\ref{eq:deltaz}.
Far from equilibrium (cf. Figs.~\ref{fig:md-ld-th}(b--c)) where $|z-z_E| > \Delta z_H = {\textstyle \frac{1}{2}} \pi \Delta U/ l$, the relaxation is exponential at the ``fast'' rate $1/T_D$ predicted by Eqs.~\ref{eq:langevin}--\ref{eq:energy_smooth} for a stable system; $T_D$ is numerically computed using large values of $K$ for which there is no metastability. 
Closer to equilibrium where $|z-z_E|< \Delta z_H$, numerical results are in close agreement with Eq.~\ref{eq:deltaz} valid for thermally-activated transitions between metastable states. 
%

%
%
%%%%%%%%%%%%%%%%%%%%%%%%%%%%%%%%%%%%%%%%%%%%%%%%%%%%%%%%%%%%%%%%%%
%  	CROSSOVERS FROM LOGARITHMIC TO EXPONENTIAL
%%%%%%%%%%%%%%%%%%%%%%%%%%%%%%%%%%%%%%%%%%%%%%%%%%%%%%%%%%%%%%%%%%
%
%%%%%%%%%%%%%%%%%%%%%%%%%%%%%%%%%%%%%%%%%%%%%%%%%%%%%%%%%%%%%%%%%%%%%%%%%%%%%%%%%%%%%%%%%%%%%%%%%%%%%%%%%%%%%%%%%%%%%%%%%%%%%%%%
% Figure3: CROSSOVERS
%%%%%%%%%%%%%%%%%%%%%%%%%%%%%%%%%%%%%%%%%%%%%%%%%%%%%%%%%%%%%%%%%%%%%%%%%%%%%%%%%%%%%%%%%%%%%%%%%%%%%%%%%%%%%%%%%%%%%%%%%%%%%%%%
\begin{figure}
\center
\includegraphics[angle=0,width=1.0\linewidth]{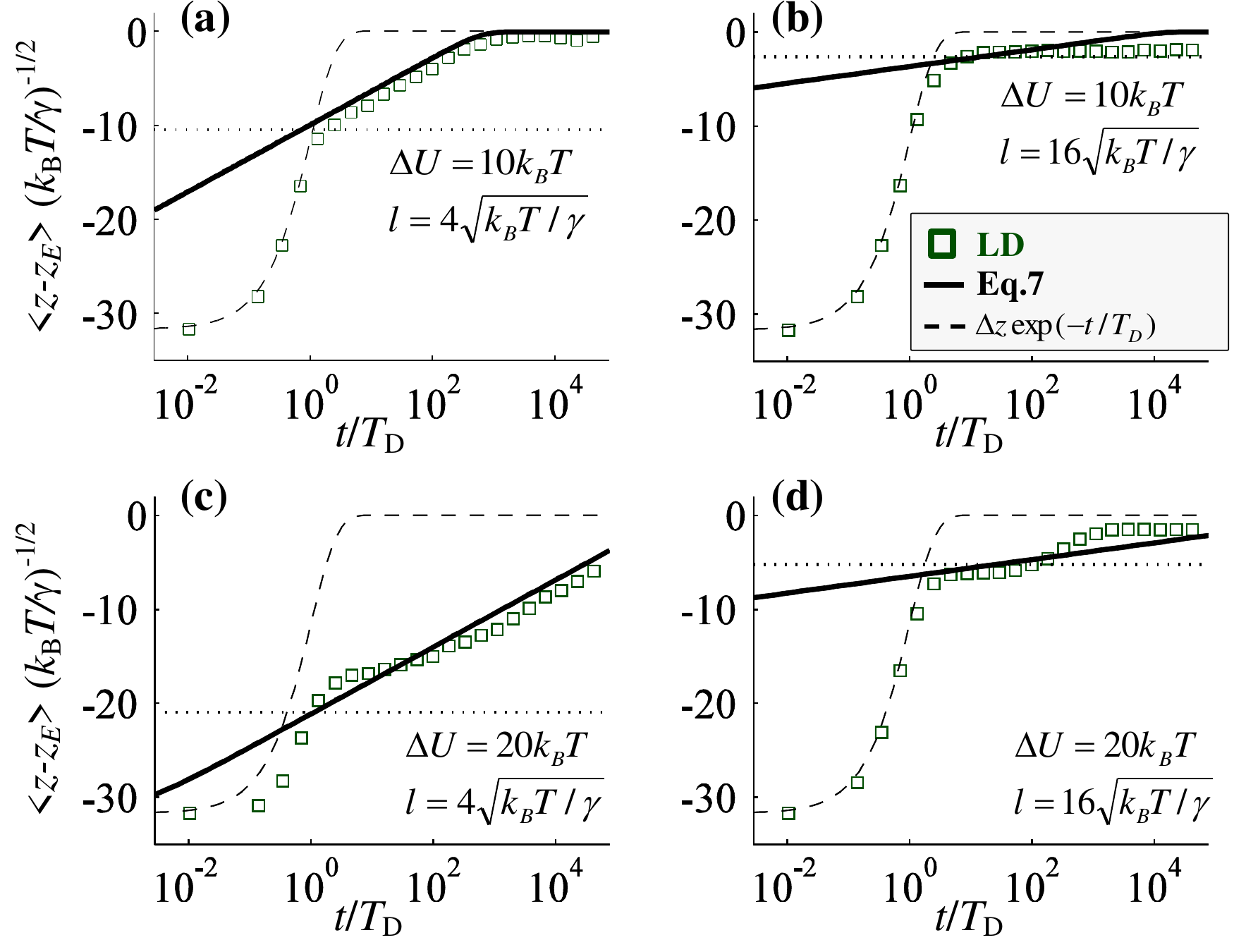}
\caption{Crossover from exponential to logarithmic relaxation. 
Solid lines indicate logarithmic relaxation given by Eq.~\ref{eq:deltaz}.
Dashed lines correspond to an exponential decay at rate $1/T_D$.
Dotted horizontal lines show the distance to equilibrium $\Delta z_H= {\textstyle \frac{1}{2}} \pi \Delta U/K l$ near which the crossover occurs.
}
\label{fig:crossover}
\end{figure}
%%%%%%%%%%%%%%%%%%%%%%%%%%%%%%%%%%%%%%%%%%%%%%%%%%%%%%%%%%%%%%%%%%%%%%%%%%%%%%%%%%%%%%%%%%%%%%%%%%%%%%%%%%%%%%%%%%%%%%%%%%%%%%%%
%
Additional simulations for larger energy barriers $\Delta U =$~10--20$k_BT$, different fluctuation wavelengths $l=$~4--16$\sqrt{k_BT/\gamma}$, and a larger separation from equilibrium $|\Delta z|=~32\sqrt{k_BT/\gamma}=$~20--82$L_H$ are reported in Fig.~\ref{fig:crossover}.
The resulting relaxation times $T_H$ in the logarithmic regime are extremely long and these simulations were only feasible via LD.
At a distance $\Delta z_H = {\textstyle \frac{1}{2}} \pi \Delta U/K l$ we observe the crossover from a fast exponential relaxation, driven by surface energy minimization, to a slow logarithmic relaxation predicted by Eq.~\ref{eq:deltaz}, driven by the thermally-activated escape from metastable states.
The crossover to a slow logarithmic relaxation is delayed, or even prevented, when increasing the wavelength $l$ of the fluctuation, cf. Figs~\ref{fig:crossover}(a--b), or decreasing the energy barrier $\Delta U$, cf. Fig.~\ref{fig:crossover}(a) and Fig.~\ref{fig:crossover}(c).

%%%%%%%%%%%%%%%%%%%%%%%%%%%%%%%%%%%%%%%%%%%%%%%%%%%%%%%%%%%%%%%%%%
%  	THE HOP TIME: DEPENDENCE ON FLUCTUATION FEATURE
%%%%%%%%%%%%%%%%%%%%%%%%%%%%%%%%%%%%%%%%%%%%%%%%%%%%%%%%%%%%%%%%%%
It is useful to examine how features of the energy fluctuations and physical properties of the media determine the relaxation time $T_H$ in the logarithmic regime.  
According to Eq.~\ref{eq:time} the dimensionless hop time $T_H/T_D$ is a function of the dimensionless amplitude and wavelength of the energy fluctuation $T_H/T_D=f(\Delta U/k_BT, l/\sqrt{k_BT/\gamma})$.
Given the functional form of Eq.~\ref{eq:time}, it is convenient to analyze the function $\log(T_H/T_D)=P-Q$, where $P=\Delta U/k_BT+{\textstyle\frac{\pi}{4}} \gamma l^2/k_BT$ is the dominant contribution.
One finds that not only increasing the energy barrier $\Delta U$, but also increasing the wavelength $l$ causes an exponential increase in the relaxation time $T_H$. 
Extremely large relaxation times $T_H$ produced by large fluctuation wavelengths $l>\sqrt{k_BT/\gamma}$, indicate that the particle will be jammed, i.e. prevented from reaching equilibrium, as soon as the hopping motion begins.
However, when the wavelength $l$ of the heterogeneity is large the particle can get much closer to equilibrium before jamming because the hopping motion begins at a much later stage when $|z-z_E|\simeq \Delta z_H \sim \Delta U/K l$.
Therefore, a logarithmic relaxation on experimentally accessible time scales can only be observed when the energy fluctuation wavelength $l<\sqrt{k_BT/\gamma}$ is smaller than the mean-square displacement of the fluid interface.

%%%%%%%%%%%%%%%%%%%%%%%%%%%%%%%%%%%%%%%%%%%%%%%%%%%%%%%%%%%%%%%%%%
%  	THE HOP TIME: CONNECTION W/EXPERIMENTAL DATA
%%%%%%%%%%%%%%%%%%%%%%%%%%%%%%%%%%%%%%%%%%%%%%%%%%%%%%%%%%%%%%%%%%
In order to discuss our findings let us consider a physical system of practical importance, a one-micron radius and nearly spherical particle adsorbed at a water-oil interface for which $\gamma=0.04$N/m and $k_BT=4\times 10^{-21}$J at room temperature.
The contact line perimeter is then $p\simeq 6 \times 10^{-6}$m.
We further consider a moderate energy barrier $\Delta U=\gamma \Delta A=30k_BT$, caused by a surface heterogeneity of area $\Delta A\simeq$ 3 nm$^2$.
If the defect is well localized only a small portion of the contact line hops over the defect while most of its perimeter remains pinned.
Hopping over localized defects, the contact line moves in steps of average length $A/p\simeq 0.5 \times 10^{-12}$m. 
This distance corresponds to an energy fluctuation wavelength $l\simeq 0.002\sqrt{k_BT/\gamma}$ which results in a relaxation time $T_H\simeq 10^{3}$s.
The logarithmic regime will then begin at $|z-z_E|\lesssim 5\times 10^{-7}$m, i.e. half a radius away from the expected equilibrium. 
This scenario seems to describe the experimental conditions studied in Ref.~\cite{kaz2011}.
Moreover, using a wavelength $l\simeq$~0.5--1$\times 10^{-12}$m and energy barriers $\Delta U\simeq$~15--30$k_BT$, we find that Eq.~\ref{eq:deltaz} fits closely the experimental data reported in Ref.~\cite{kaz2011}.
%

%
% Conclusions
%
In conclusion, we have derived an analytical expression for the adsorption dynamics of a colloidal particle in the case of thermally-activated relaxation to equilibrium.
We found that moderate energy barriers, of order 10$k_BT$, due to localized nanoscale heterogeneities can give rise to physical aging as the particle approaches the expected equilibrium position $z_E$. 
Numerical simulations show a crossover from a fast exponential relaxation to a slow logarithmic relaxation at a distance $|z-z_E|\sim \Delta U/ \gamma l$ from the expected equilibrium.
These results demonstrate a nontrivial relaxation dynamics of adsorption of colloidal particles at a fluid interface.

%%%%%%%%%%%%%%%%%%%%%%%%%%%%%%%%%%%%%%%%%%%%%%%%%%%%%%%%%%%%%%
%%%%%%%%%%%%%%%%%%%%%%%%%%%%%%%%%%%%%%%%%%%%%%%%%%%%%%%%%%%%%%
% ACKNOWLEDGEMENTS
%%%%%%%%%%%%%%%%%%%%%%%%%%%%%%%%%%%%%%%%%%%%%%%%%%%%%%%%%%%%%%
%%%%%%%%%%%%%%%%%%%%%%%%%%%%%%%%%%%%%%%%%%%%%%%%%%%%%%%%%%%%%%
%
We are thankful to Prof. Vinothan N. Manoharan and Ms. Anna Wang for helpful discussions.
This work was supported by the NSF PREM (DMR-0934206).
%

%
%
%\newpage 
%\bibliography{adsorption}

\begin{thebibliography}{10}%
\makeatletter
\providecommand \@ifxundefined [1]{%
 \ifx #1\undefined \expandafter \@firstoftwo
 \else \expandafter \@secondoftwo
\fi
}%
\providecommand \@ifnum [1]{%
 \ifnum #1\expandafter \@firstoftwo
 \else \expandafter \@secondoftwo
\fi
}%
\providecommand \enquote [1]{``#1''}%
\providecommand \bibnamefont  [1]{#1}%
\providecommand \bibfnamefont [1]{#1}%
\providecommand \citenamefont [1]{#1}%
\providecommand\href[0]{\@sanitize\@href}%
\providecommand\@href[1]{\endgroup\@@startlink{#1}\endgroup\@@href}%
\providecommand\@@href[1]{#1\@@endlink}%
\providecommand \@sanitize [0]{\begingroup\catcode`\&12\catcode`\#12\relax}%
\@ifxundefined \pdfoutput {\@firstoftwo}{%
 \@ifnum{\z@=\pdfoutput}{\@firstoftwo}{\@secondoftwo}%
}{%
 \providecommand\@@startlink[1]{\leavevmode}%
 \providecommand\@@endlink[0]{}%
}{%
 \providecommand\@@startlink[1]{%
  \leavevmode
  \pdfstartlink
   attr{/Border[0 0 1 ]/H/I/C[0 1 1]}%
   user{/Subtype/Link/A<</Type/Action/S/URI/URI(#1)>>}%
  \relax
 }%
 \providecommand\@@endlink[0]{\pdfendlink}%
}%
\providecommand \url  [0]{\begingroup\@sanitize \@url }%
\providecommand \@url [1]{\endgroup\@href {#1}{\urlprefix}}%
\providecommand \urlprefix [0]{URL }%
\providecommand \Eprint[0]{\href }%
\@ifxundefined \urlstyle {%
  \providecommand \doi [1]{doi:\discretionary{}{}{}#1}%
}{%
  \providecommand \doi [0]{doi:\discretionary{}{}{}\begingroup
  \urlstyle{rm}\Url }%
}%
\providecommand \doibase [0]{http://dx.doi.org/}%
\providecommand \Doi[1]{\href{\doibase#1}}%
\providecommand \bibAnnote [3]{%
  \BibitemShut{#1}%
  \begin{quotation}\noindent
    \textsc{Key:}\ #2\\\textsc{Annotation:}\ #3%
  \end{quotation}%
}%
\providecommand \bibAnnoteFile [2]{%
  \IfFileExists{#2}{\bibAnnote {#1} {#2} {\input{#2}}}{}%
}%
\providecommand \typeout [0]{\immediate \write \m@ne }%
\providecommand \selectlanguage [0]{\@gobble}%
\providecommand \bibinfo [0]{\@secondoftwo}%
\providecommand \bibfield [0]{\@secondoftwo}%
\providecommand \translation [1]{[#1]}%
\providecommand \BibitemOpen[0]{}%
\providecommand \bibitemStop [0]{}%
\providecommand \bibitemNoStop [0]{.\EOS\space}%
\providecommand \EOS [0]{\spacefactor3000\relax}%
\providecommand \BibitemShut [1]{\csname bibitem#1\endcsname}%
%</preamble>
\bibitem{particles}%
  \BibitemOpen
  \bibfield{author}{%
  \bibinfo {author} {\bibfnamefont{P.}~\bibnamefont{Kralchevsky}}\ and\
  \bibinfo {author} {\bibfnamefont{K.}~\bibnamefont{Nagayama}},\ }%
  \emph{\bibinfo {title} {Particles at Fluids Interfaces and Membranes}},\
  Vol.~\bibinfo {volume} {10}\ (\bibinfo {publisher} {Elsevier Science},\
  \bibinfo {year} {2001})%
  \bibAnnoteFile{NoStop}{particles}%
\bibitem{velev2009}%
  \BibitemOpen
  \bibfield{author}{%
  \bibinfo {author} {\bibfnamefont{O.}~\bibnamefont{Velev}}\ and\ \bibinfo
  {author} {\bibfnamefont{S.}~\bibnamefont{Gupta}},\ }%
  \bibfield{journal}{%
  \bibinfo {journal} {Adv. Mater.}\ }%
  \textbf{\bibinfo {volume} {21}},\ \bibinfo {pages} {1897} (\bibinfo {year}
  {2009})%
  \bibAnnoteFile{NoStop}{velev2009}%
\bibitem{goesmann2010}%
  \BibitemOpen
  \bibfield{author}{%
  \bibinfo {author} {\bibfnamefont{H.}~\bibnamefont{Goesmann}}\ and\ \bibinfo
  {author} {\bibfnamefont{C.}~\bibnamefont{Feldmann}},\ }%
  \bibfield{journal}{%
  \bibinfo {journal} {Angew. Chem. Int. Ed.}\ }%
  \textbf{\bibinfo {volume} {49}},\ \bibinfo {pages} {1362} (\bibinfo {year}
  {2010})%
  \bibAnnoteFile{NoStop}{goesmann2010}%
\bibitem{mezzenga2005}%
  \BibitemOpen
  \bibfield{author}{%
  \bibinfo {author} {\bibfnamefont{R.}~\bibnamefont{Mezzenga}}, \bibinfo
  {author} {\bibfnamefont{P.}~\bibnamefont{Schurtenberger}}, \bibinfo {author}
  {\bibfnamefont{A.}~\bibnamefont{Burbidge}},\ and\ \bibinfo {author}
  {\bibfnamefont{M.}~\bibnamefont{Michel}},\ }%
  \bibfield{journal}{%
  \bibinfo {journal} {Nat. Mater.}\ }%
  \textbf{\bibinfo {volume} {4}},\ \bibinfo {pages} {729} (\bibinfo {year}
  {2005})%
  \bibAnnoteFile{NoStop}{mezzenga2005}%
\bibitem{dinsmore2002}%
  \BibitemOpen
  \bibfield{author}{%
  \bibinfo {author} {\bibfnamefont{A.}~\bibnamefont{Dinsmore}}, \bibinfo
  {author} {\bibfnamefont{M.}~\bibnamefont{Hsu}}, \bibinfo {author}
  {\bibfnamefont{M.}~\bibnamefont{Nikolaides}}, \bibinfo {author}
  {\bibfnamefont{M.}~\bibnamefont{Marquez}}, \bibinfo {author}
  {\bibfnamefont{A.}~\bibnamefont{Bausch}},\ and\ \bibinfo {author}
  {\bibfnamefont{D.}~\bibnamefont{Weitz}},\ }%
  \bibfield{journal}{%
  \bibinfo {journal} {Science}\ }%
  \textbf{\bibinfo {volume} {298}},\ \bibinfo {pages} {1006} (\bibinfo {year}
  {2002})%
  \bibAnnoteFile{NoStop}{dinsmore2002}%
\bibitem{velikov2008}%
  \BibitemOpen
  \bibfield{author}{%
  \bibinfo {author} {\bibfnamefont{K.}~\bibnamefont{Velikov}}\ and\ \bibinfo
  {author} {\bibfnamefont{E.}~\bibnamefont{Pelan}},\ }%
  \bibfield{journal}{%
  \bibinfo {journal} {Soft Matter}\ }%
  \textbf{\bibinfo {volume} {4}},\ \bibinfo {pages} {1964} (\bibinfo {year}
  {2008})%
  \bibAnnoteFile{NoStop}{velikov2008}%
\bibitem{pieranski1980}%
  \BibitemOpen
  \bibfield{author}{%
  \bibinfo {author} {\bibfnamefont{P.}~\bibnamefont{Pieranski}},\ }%
  \bibfield{journal}{%
  \bibinfo {journal} {Phys. Rev. Lett.}\ }%
  \textbf{\bibinfo {volume} {45}},\ \bibinfo {pages} {569} (\bibinfo {year}
  {1980})%
  \bibAnnoteFile{NoStop}{pieranski1980}%
\bibitem{israelachvili1989}%
  \BibitemOpen
  \bibfield{author}{%
  \bibinfo {author} {\bibfnamefont{J.}~\bibnamefont{Israelachvili}}\ and\
  \bibinfo {author} {\bibfnamefont{M.}~\bibnamefont{Gee}},\ }%
  \bibfield{journal}{%
  \bibinfo {journal} {Lang.}\ }%
  \textbf{\bibinfo {volume} {5}},\ \bibinfo {pages} {288} (\bibinfo {year}
  {1989})%
  \bibAnnoteFile{NoStop}{israelachvili1989}%
\bibitem{bormashenko2012}%
  \BibitemOpen
  \bibfield{author}{%
  \bibinfo {author} {\bibfnamefont{E.}~\bibnamefont{Bormashenko}},\ }%
  \bibfield{journal}{%
  \bibinfo {journal} {Colloid Polym. Sci.},\ \bibinfo {pages} {1}}%
   (\bibinfo {year} {2012})%
  \bibAnnoteFile{NoStop}{bormashenko2012}%
\bibitem{paunov2003}%
  \BibitemOpen
  \bibfield{author}{%
  \bibinfo {author} {\bibfnamefont{V.}~\bibnamefont{Paunov}},\ }%
  \bibfield{journal}{%
  \bibinfo {journal} {Lang.}\ }%
  \textbf{\bibinfo {volume} {19}},\ \bibinfo {pages} {7970} (\bibinfo {year}
  {2003})%
  \bibAnnoteFile{NoStop}{paunov2003}%
\bibitem{arnaudov2009}%
  \BibitemOpen
  \bibfield{author}{%
  \bibinfo {author} {\bibfnamefont{L.}~\bibnamefont{Arnaudov}}, \bibinfo
  {author} {\bibfnamefont{O.}~\bibnamefont{Cayre}}, \bibinfo {author}
  {\bibfnamefont{M.}~\bibnamefont{Stuart}}, \bibinfo {author}
  {\bibfnamefont{S.}~\bibnamefont{Stoyanov}},\ and\ \bibinfo {author}
  {\bibfnamefont{V.}~\bibnamefont{Paunov}},\ }%
  \bibfield{journal}{%
  \bibinfo {journal} {Phys. Chem. Chem. Phys.}\ }%
  \textbf{\bibinfo {volume} {12}},\ \bibinfo {pages} {328} (\bibinfo {year}
  {2009})%
  \bibAnnoteFile{NoStop}{arnaudov2009}%
\bibitem{stocco2011}%
  \BibitemOpen
  \bibfield{author}{%
  \bibinfo {author} {\bibfnamefont{A.}~\bibnamefont{Stocco}}, \bibinfo {author}
  {\bibfnamefont{E.}~\bibnamefont{Rio}}, \bibinfo {author}
  {\bibfnamefont{B.}~\bibnamefont{Binks}},\ and\ \bibinfo {author}
  {\bibfnamefont{D.}~\bibnamefont{Langevin}},\ }%
  \bibfield{journal}{%
  \bibinfo {journal} {Soft Matter}\ }%
  \textbf{\bibinfo {volume} {7}},\ \bibinfo {pages} {1260} (\bibinfo {year}
  {2011})%
  \bibAnnoteFile{NoStop}{stocco2011}%
\bibitem{du2010}%
  \BibitemOpen
  \bibfield{author}{%
  \bibinfo {author} {\bibfnamefont{K.}~\bibnamefont{Du}}, \bibinfo {author}
  {\bibfnamefont{E.}~\bibnamefont{Glogowski}}, \bibinfo {author}
  {\bibfnamefont{T.}~\bibnamefont{Emrick}}, \bibinfo {author}
  {\bibfnamefont{T.}~\bibnamefont{Russell}},\ and\ \bibinfo {author}
  {\bibfnamefont{A.}~\bibnamefont{Dinsmore}},\ }%
  \bibfield{journal}{%
  \bibinfo {journal} {Lang.}\ }%
  \textbf{\bibinfo {volume} {26}},\ \bibinfo {pages} {12518} (\bibinfo {year}
  {2010})%
  \bibAnnoteFile{NoStop}{du2010}%
\bibitem{kaz2011}%
  \BibitemOpen
  \bibfield{author}{%
  \bibinfo {author} {\bibfnamefont{D.}~\bibnamefont{Kaz}}, \bibinfo {author}
  {\bibfnamefont{R.}~\bibnamefont{McGorty}}, \bibinfo {author}
  {\bibfnamefont{M.}~\bibnamefont{Mani}}, \bibinfo {author}
  {\bibfnamefont{M.}~\bibnamefont{Brenner}},\ and\ \bibinfo {author}
  {\bibfnamefont{V.}~\bibnamefont{Manoharan}},\ }%
  \bibfield{journal}{%
  \bibinfo {journal} {Nat. Mater.}\ }%
  \textbf{\bibinfo {volume} {11}},\ \bibinfo {pages} {138} (\bibinfo {year}
  {2012})%
  \bibAnnoteFile{NoStop}{kaz2011}%
\bibitem{Strum1977}%
  \BibitemOpen
  \bibfield{author}{%
  \bibinfo {author} {\bibfnamefont{L.~C.~E.}\ \bibnamefont{Strum}},\ }%
  \bibfield{journal}{%
  \bibinfo {journal} {Polym. Eng. Sci.}\ }%
  \textbf{\bibinfo {volume} {17}},\ \bibinfo {pages} {165} (\bibinfo {year}
  {1977})%
  \bibAnnoteFile{NoStop}{Strum1977}%
\bibitem{aveyard1996}%
  \BibitemOpen
  \bibfield{author}{%
  \bibinfo {author} {\bibfnamefont{R.}~\bibnamefont{Aveyard}}\ and\ \bibinfo
  {author} {\bibfnamefont{J.}~\bibnamefont{Clint}},\ }%
  \bibfield{journal}{%
  \bibinfo {journal} {Faraday Trans.}\ }%
  \textbf{\bibinfo {volume} {92}},\ \bibinfo {pages} {85} (\bibinfo {year}
  {1996})%
  \bibAnnoteFile{NoStop}{aveyard1996}%
\bibitem{marmur1998}%
  \BibitemOpen
  \bibfield{author}{%
  \bibinfo {author} {\bibfnamefont{A.}~\bibnamefont{Marmur}},\ }%
  \bibfield{journal}{%
  \bibinfo {journal} {Colloid. Surf. A}\ }%
  \textbf{\bibinfo {volume} {136}},\ \bibinfo {pages} {81} (\bibinfo {year}
  {1998})%
  \bibAnnoteFile{NoStop}{marmur1998}%
\bibitem{Note1}%
  \BibitemOpen
  \bibinfo {note} {MD simulations required extremely long runs (i.e. over 10
  million timesteps) and several (7--10) independent realizations; running in
  parallel (7 CPUs at 2.20GHz) the simulations required over 300 hours of CPU
  time.}%
  \bibAnnoteFile{Stop}{Note1}%
\bibitem{kramers}%
  \BibitemOpen
  \bibfield{author}{%
  \bibinfo {author} {\bibfnamefont{H.}~\bibnamefont{Kramers}},\ }%
  \bibfield{journal}{%
  \bibinfo {journal} {Physica}\ }%
  \textbf{\bibinfo {volume} {7}},\ \bibinfo {pages} {284} (\bibinfo {year}
  {1940})%
  \bibAnnoteFile{NoStop}{kramers}%
\bibitem{hanggi1986}%
  \BibitemOpen
  \bibfield{author}{%
  \bibinfo {author} {\bibfnamefont{P.}~\bibnamefont{Hanggi}},\ }%
  \bibfield{journal}{%
  \bibinfo {journal} {Journal of Statistical Physics}\ }%
  \textbf{\bibinfo {volume} {42}},\ \bibinfo {pages} {105} (\bibinfo {year}
  {1986})%
  \bibAnnoteFile{NoStop}{hanggi1986}%
\bibitem{rapaport}%
  \BibitemOpen
  \bibfield{author}{%
  \bibinfo {author} {\bibfnamefont{D.}~\bibnamefont{Rapaport}},\ }%
  \emph{\bibinfo {title} {The Art of Molecular Dynamics Simulation}},\ \bibinfo
  {edition} {2nd}\ ed.\ (\bibinfo {publisher} {Cambridge University, New
  York},\ \bibinfo {year} {1995})%
  \bibAnnoteFile{NoStop}{rapaport}%
\bibitem{koplik2006}%
  \BibitemOpen
  \bibfield{author}{%
  \bibinfo {author} {\bibfnamefont{J.}~\bibnamefont{Koplik}}\ and\ \bibinfo
  {author} {\bibfnamefont{J.~R.}\ \bibnamefont{Banavar}},\ }%
  \bibfield{journal}{%
  \bibinfo {journal} {Phys. Rev. Lett.}\ }%
  \textbf{\bibinfo {volume} {96}},\ \bibinfo {pages} {044505} (\bibinfo {year}
  {2006})%
  \bibAnnoteFile{NoStop}{koplik2006}%
\bibitem{drazer2002}%
  \BibitemOpen
  \bibfield{author}{%
  \bibinfo {author} {\bibfnamefont{G.}~\bibnamefont{Drazer}}, \bibinfo {author}
  {\bibfnamefont{J.}~\bibnamefont{Koplik}}, \bibinfo {author}
  {\bibfnamefont{A.}~\bibnamefont{Acrivos}},\ and\ \bibinfo {author}
  {\bibfnamefont{B.}~\bibnamefont{Khusid}},\ }%
  \bibfield{journal}{%
  \bibinfo {journal} {Phys. Rev. Lett.}\ }%
  \textbf{\bibinfo {volume} {89}},\ \bibinfo {pages} {244501} (\bibinfo {year}
  {2002})%
  \bibAnnoteFile{NoStop}{drazer2002}%
\bibitem{koplik1995}%
  \BibitemOpen
  \bibfield{author}{%
  \bibinfo {author} {\bibfnamefont{J.}~\bibnamefont{Koplik}}\ and\ \bibinfo
  {author} {\bibfnamefont{J.}~\bibnamefont{Banavar}},\ }%
  \bibfield{journal}{%
  \bibinfo {journal} {Annu. Rev. Fluid Mech.}\ }%
  \textbf{\bibinfo {volume} {27}},\ \bibinfo {pages} {257} (\bibinfo {year}
  {1995})%
  \bibAnnoteFile{NoStop}{koplik1995}%
\end{thebibliography}
%
%Merlin.mbs v4.21 2009-07-09.
%

\end{document}